\input psfig
\def\puncspace{\ifmmode\,\else{\ifcat.\C{\if.\C\else\if,\C\else\if?\C\else%
\if:\C\else\if;\C\else\if-\C\else\if)\C\else\if/\C\else\if]\C\else\if'\C%
\else\space\fi\fi\fi\fi\fi\fi\fi\fi\fi\fi}%
\else\if\empty\C\else\if\space\C\else\space\fi\fi\fi}\fi}
\def\SP{\let\\=\empty\futurelet\C\puncspace}
\def\iras{{\it IRAS}\SP}
\def\dec{$\delta$\SP}

\def\maior {$\geq$\SP}
\def\kms{km~s$^{-1}$\SP}

\def\degree{$^\circ$\SP}
\def\h1{$h^{-1}$\SP}

\def\etal{{\it et al.\/}\ }
\def\ie{{\it i.e.\/}\rm,\ }
\def\eg{{\it e.g.\/}\rm,\ }
\def\lsim{~\rlap{$<$}{\lower 1.0ex\hbox{$\sim$}}}
\def\gsim{~\rlap{$>$}{\lower 1.0ex\hbox{$\sim$}}}
\def\void#1{{}}

\documentstyle[onecolumn]{mn}
\def\puncspace{\ifmmode\,\else{\ifcat.\C{\if.\C\else\if,\C\else\if?\C\else%
\if:\C\else\if;\C\else\if-\C\else\if)\C\else\if/\C\else\if]\C\else\if'\C%
\else\space\fi\fi\fi\fi\fi\fi\fi\fi\fi\fi}%
\else\if\empty\C\else\if\space\C\else\space\fi\fi\fi}\fi}
\def\SP{\let\\=\empty\futurelet\C\puncspace }
\def\etal{et\SP al.\SP }

\def\h-1{$h^{-1}$}
\begin{document}

\title[Comparison of the SFI Peculiar Velocities
with the  IRAS 1.2 Jy Gravity Field]
{Comparison of the SFI Peculiar Velocities
with the  IRAS 1.2 Jy Gravity Field}

\author [da Costa et. al.] {Luiz N. da Costa$^1$, Adi Nusser$^2$,
Wolfram Freudling$^3$, Riccardo Giovanelli$^4$, \newauthor Martha P. Haynes$^4$, John
J. Salzer$^5$
and Gary Wegner$^6$\\
$^1$European Southern Observatory, Karl-Schwarzschild Str. 2, D--85748
Garching b. M\"unchen, Germany\\
$^2$Max-Planck-Institut f\"ur Astrophysik, Karl-Schwarzschild-Str. 1,  D--85749
Garching b. M\"unchen, Germany\\
$^3$Space Telescope--European Coordinating Facility, European
Southern Observatory, Karl-Schwarzschild-Str. 2, D--85748
Garching b. M\"unchen, Germany\\
$^4$Center for Radiophysics and Space Research
and National Astronomy and Ionosphere Center,
Cornell University, Ithaca, NY 14953\\
$^5$Dept. of Astronomy, Wesleyan University, Middletown, CT 06457\\
$^6$Dept. of Physics and Astronomy, Dartmouth College, Hanover, 
NH 03755}

\maketitle

\begin{abstract}

We present a comparison between the peculiar velocity fields measured
from the SFI all-sky Sbc-Sc Tully-Fisher catalog and that derived
from the \iras 1.2 Jy redshift survey galaxy distribution. The
analysis is based on the expansion of these data
in redshift space using smooth orthonormal functions and is
performed using low and high resolution expansions, with an effective
smoothing scale which increases almost linearly with redshift. The
effective smoothing scales at 3000 \kms are 1500\kms and 1000\kms for
the low and high resolution filters.  The agreement
between the high and low resolution SFI velocity maps is excellent.
The general features in the filtered SFI and \iras velocity fields
agree remarkably well within 6000\kms.  This good agreement between
the fields allows us to determine the parameter
$\beta=\Omega^{0.6}/b$, where $\Omega $ is the cosmological density
parameter and $b$ is the linear biasing factor.  From a likelihood
analysis on the SFI and \iras modes we find that $\beta=0.6\pm 0.1$
independently of the resolution of the modal expansion.  For this
value of $\beta$, the residual fields for the two filters show no
systematic variations within 6000 \kms.  Most remarkable is the
lack of any coherent, redshift dependent dipole flow in the residual
field.

\end{abstract}

\begin{keywords}
cosmology: observations -- dark matter -- large scale structure of Universe
\end{keywords}

\section {Introduction}
Measurements of the peculiar motions of galaxies in the nearby universe
represent one of the most powerful tools currently available to probe
mass fluctuations on large scales ($\lsim 100 {\rm h^{-1}} Mpc$).
Furthermore, in
conjunction with redshift surveys, the relationship
between the distribution of luminous and dark matter
can be investigated. Peculiar motions
also offer an alternative estimate of the cosmological density
parameter $\Omega$ on intermediate scales, complementing other dynamical
measures on smaller scales (Fisher \etal 1994, Carlberg \etal 1996), and
global determinations such as the magnitude-redshift diagram of SN Ia (\eg
Perlmutter \etal 1997) and statistics of gravitational lenses (Kochanek
1996). Estimates of $\Omega$ from peculiar motions can be derived either
using general assumptions about the initial fluctuations (\eg Nusser \&
Dekel 1993) that are independent of the galaxy distribution, or by combining
the peculiar velocity data with redshift surveys (e.g. Dekel \etal 1993,
Davis, Nusser \& Willick 1996, Freudling \etal 1997).  The latter case
requires an assumption about the biasing relation between the galaxy and
the dark matter distribution, which is usually taken to be linear, and
leads to an estimate of the parameter $\beta = \Omega^{0.6}/b$ where $b$
is the biasing factor. 

Work on this subject has led to some puzzling results.
Primarily this is because
most earlier work was based in a relatively sparse and
inhomogeneuous set of galaxies with measured peculiar velocities. 
Recently, the observational situation has dramatically improved with the
completion of large redshift-distance samples in both hemispheres. The
most notable are the Mathewson, Ford \& Buchhorn survey (1992, MFB92)
with about 1200 galaxies with I-band photometry and measured rotational
velocities, either from radio observations of 21-cm linewidths or
optical rotation curves, and the I-band Tully-Fisher (TF) distance
survey of about 2000 spiral galaxies in the field (SFI survey, da Costa
\etal 1996, Haynes \etal 1997) and in the direction of 24 clusters (SCI
survey, Giovanelli \etal 1997a). These TF surveys have been used to
construct two largely independent all-sky redshift-distance catalogs:
the Mark~III catalog, which is a compilation of all available data to
date (\eg Willick \etal 1997); the SFI catalog (\eg da Costa \etal 1996,
Haynes \etal 1997) which combines a pruned version of the MFB92 data,
south of $\delta = -45^\circ$, with the SFI survey (\eg da Costa \etal
1996, Haynes \etal 1997).

The SFI catalog has recently been used 
by da Costa \etal (1996) to reconstruct the mass density
and three-dimensional velocity fields.
Significant differences were found relative to earlier
reconstructions.  In particular, for the first time the measured
velocity field showed a bifurcation with some galaxies flowing towards
the Great Attractor (GA) while others moving towards the
Perseus-Pisces complex (PP), a feature seen in the
\iras velocity field. In that paper it was conjectured that the better
agreement between the gross features of the flows was likely to have
an impact on the accuracy with which the parameter $\beta$ could be
determined since it relies on a good match between the observed and
predicted velocity fields.

The main goals of the present paper are: 1) to investigate more
quantitatively the agreement between the measured radial peculiar
velocities and the \iras predicted gravity field; 2) to use the
velocity-velocity comparison to determine the parameter $\beta$; 3) to
compare the results with those obtained from a similar analysis of the
Mark~III catalog.

The current analysis uses the  the method of orthogonal
mode-expansion (ITF method) developed by Nusser \& Davis (1995, ND95). The
main advantages of the method are that it uses the inverse TF relation,
which as shown by Schechter (1980) reduces the effects of Malmquist
bias, requires no binning of the data and provides a smooth map of the
velocity field. The method is ideal for comparing different datasets
and provides a useful way of displaying the velocity field. Furthermore,
since it has recently been applied to the Mark~III catalog by Davis,
Nusser \& Willick (1996, DNW) we can indirectly compare these
two catalogs. Understanding the differences between these catalogs is
important for this may make it possible to
combine all the currently available samples.

In section 2, we briefly describe the SFI redshift-distance catalog.
In section 3, we review the basics of the ITF method and describe our
choice of basis function.  In section 4, we apply the method to the
SFI catalog and present low and high resolution maps for the SFI
radial velocity field and compare them to the predicted \iras gravity
field for different assumed values of $\beta$. The likelihood method
used to derive $\beta$ is discussed in section 4. A brief summary of
our conclusions is presented in section 5.

\section {Data}

The TF data used here consists of two sets: the first is the SFI
survey, an I-band and 21 cm survey of Sbc-Sc galaxies with
inclinations $\gsim$ 45\degree, north of \dec
\maior -45\degree and galactic latitudes $b > 10^{\circ}$. Galaxies
were drawn according to the following redshift-dependent criteria:
$2.5' < a < 5' \ cz < $ 3000 \kms; $1.6' < a < 5' \ 3000 < cz < $ 5000
\kms;$1.3' < a < 5' \ 5000 < cz < $ 7500 \kms . 
The second is a pruned version of the MFB92 survey, including only
Sbc-Sc galaxies. The original MFB92 measurements of magnitude and
rotational velocities, either from 21 cm line widths or optical
rotation curves, were converted into the SFI system using galaxies in
common with the SFI survey.  For the MFB92 galaxies with only
optically measured line widths, we have used the kinematically
centered rotation curves of Persic \& Salucci (1995) to derive the
rotational velocities at one optical radius in order to bring them on
a common scale. These values were used to calibrate the relation
between the MFB92 rotational velocities and our 21 cm line widths. We
also use the overlapping set of galaxies to transform the MFB92
magnitudes into the SFI system (Giovanelli \etal 1997a).

\section {Method}

In linear theory, the velocity field is related to gravity by
the parameter $\beta = \Omega^{0.6}/b$. Our aim is to 
estimate of the underlying peculiar velocity field from the SFI data
and compare it with the velocity field predicted from the gravity field 
computed from the \iras galaxy redshift distribution.

In order to derive a smooth velocity field from the SFI data we use
the method developed by ND95, based on the
inverse TF relation.  We assume that the rotational velocity parameter
$\eta = log(W) -2.5$, where $W$ is the line width, is related to its
absolute magnitude $M$ by means of a linear inverse TF relation

\begin{equation}
\eta = \gamma M +\eta_o ,
\end{equation}
where  $\gamma$ and $\eta_0$ are, respectively, the slope and the
zero point of the relation. 

The method is designed to describe the underlying velocity field by a
set of smooth functions.  Following ND95 we write $M_i = M_{0i} +
P_i$, where $M_{0i} = m -5\log(s_i) -15 $, $P_i=5 -\log (1 - u_i/s_i)$,
$m_i$ is the apparent magnitude of the galaxy, $s_i=cz_i$ is its
redshift in \kms
and $u_i$ its radial peculiar velocity. The method assumes that the
function $P$ can be expanded in a set of smooth functions,
\ie .
\begin{equation}
P_i = \sum \alpha^j  F_i^j ,
\end{equation}
where  the functions $F_i^j$ are orthonormal 
in the space of the data points, \ie 
$\sum_i  F_i^j F_i^{j^\prime} = \delta_K^{j {j^\prime}}$.

 The coefficients $\alpha^j$ and the ITF parameters, $\gamma$ and
 $\eta_0$, are then found simultaneously by minimizing
\begin{equation}
\chi^2 = 
\sum { [\eta_0 +\gamma ( M_{0i}+\sum_j \alpha^j F_i^j) - \eta_i]^2 \over \sigma^2_\eta} ,
\end{equation}
where $\sigma_{\eta}$ is the rms scatter of the inverse relation.

We choose the zeroth mode to describe a Hubble flow in the space of
the data set, i.e, $F_i^0=1/\sqrt{N_g}$ where $N_g$ is the
number of galaxies in the sample. Therefore the mode $\alpha^0$ fixes
the zero point of the ITF relation. Here we arbitrarily set
$\alpha^0=0$ in the following analysis.  In the comparison between the
\iras and ITF fields we subsequently remove such a Hubble flow from
the
\iras velocity field. Thanks to the uniformity of the
SFI sample, this correction is negligible. 
Following DNW we construct the
higher order functions $F_j$ from spherical harmonics $Y_l^m$
for the angular wavefunctions (Fisher \etal 1995) and derivatives of
spherical Bessel functions, $j_l[y(z)]$ for the radial basis
functions, where the transformation from $z$ to $y$ in the argument of
these Bessel functions is designed to make them oscillate non-uniformly
with depth in order to match the spatial distribution of the TF data.
The use of the coordinate $y$ instead of $z$ significantly reduces the
number of parameters necessary to fit the underlying velocity
field in terms of our velocity model (DNW).  For the reasons given in
DNW, we formulate our model to describe the velocity field with
respect to the motion of the Local Group.  The velocity model can be
written in the form
\begin{equation}
P(s,\theta,\phi)=\sum_{n=0}^{n_{max}}\sum_{l=0}^{l_{max}}\sum_{m=-l}^{m=l}
{a_{nlm} \over s}\left[j^{\prime}_{l}\left(k_ny(s)\right)-c_{l1}\right]
Y_{lm}\left(\theta,\phi\right) \quad . \label{pexp}
\end{equation}
The constant $c_{l1}$ is non-zero only for the dipole term and
is introduced to ensure that $P=0$ at the origin since we work 
in the LG frame. 
ND95 give details of the derivation of the orthonormal functions $F_i^j$. 
For the SFI data, we find that  the following transformation:
\begin{equation}
y=\frac{s}{1000} \left[ 1+\left(\frac{s}{1000}\right)^2\right]^{-1/2} ,
\label{ytrans}
\end{equation}
yields a better fit than $y=s$. For instance, for an assumed value of
$\sigma_{\eta}=0.065$ (see discussion below), when the flow of 1114
SFI galaxies within 6000~\kms was fitted with 40 modes, the $\chi^2$
was reduced from 1292 to 1074 with the transformation (\ref{ytrans})
and to 1092 with $y=s$.

\section {Analysis of the SFI and {\it IRAS} Velocity Fields}

We apply the above method to the $1114$ galaxies
within a redshift of 6000~\kms in the SFI sample.  We will present
results for low and high resolution smoothings which we label LR and
HR, respectively.  The LR smoothing corresponds to
$l_{max}=3,n_{max}=3$ which requires 40 modes while the HR smoothing
is obtained from expanding the peculiar velocity fields with 74 modes
corresponding to $l_{max}=4,n_{max}=4$. In figure 1 we show the
variation of the radial resolution as a function of the redshift for
the two filters.

The inverse TF parameters we obtain after the modal expansion are,
($\gamma,\eta_0=-0.117,-2.47$) and ($\gamma , \eta_0=-0.118,-2.47$)
for the LR and HR filters, respectively.  For $\sigma_{\eta}=0.0656$
the $\chi^2$ of the ITF regression of observed versus predicted
line-widths drops from 1292 to 1074 and 1050 for LR and HR,
respectively.  In the following analysis we work with a value of
$\sigma_\eta=0.065$.  This value yields a reduced $\chi^2$ equal to
unity for the LR filter and very close to unity for the HR filter.
Note that a significantly amaller value 
($\sigma_\eta=0.046$),
is obtained from the SCI galaxy cluster (Giovanelli
\etal 1997b) sample.
However, a direct comparison between the field and cluster samples is
not trivial. For instance, one possible reason for this difference
could be that spirals in clusters form a more homogenous population
than those in the field. Note, however, that although the scatter differs
significantly, the slopes of inverse TF relation for the SFI and SCI
differ only by 4\%.  In our analysis we do not take into account any
possible dependence of the scatter on the absolute magnitudes.  In
fact, after fitting our velocity model to the SFI data, we did not
find any evidence for magnitude dependent $\eta$ scatter.  We point
out however that the scatter is not strictly Gaussian.

In order to inspect the performance of the modal
expansion, in more detail,
in figure 2 we plot the correlation functions of the
residuals, $\Delta \eta_i=\eta_i-\gamma M_{0i}-\eta_0$, before and
after the modal expansion.  For both the LR and HR smoothings, the
amplitude of the correlation functions is consistent with zero even on
separations smaller than the resolution limit.  This plot demonstrates
that our procedure successfully extracted the signal from the TF
data.

Given an assumed value for $\beta$ we apply the method of Nusser \&
Davis (1994) to generate maps of velocity fields from the distribution
of the \iras galaxies in space.  This method generates velocity fields
non-iteratively in redshift space.  As input to this method a density
field is provided by smoothing the \iras galaxy distribution with a
gaussian window of width equal to half the mean particle separation at
a given redshift.  Because of our choice of the zeroth point for the
ITF relation, we subtract from the \iras velocity fields, any Hubble
flow like component at the position of the SFI galaxies.  The \iras
fields are then expanded in the same orthogonal set of basis functions
as employed for 
the SFI velocities.
The \iras and SFI velocities are guaranteed to have the
same resolution because the original smoothing of the \iras
density field is small compared to the resolution of the modal
expansion,

\subsection {Smooth Velocity Maps}

The resulting LR and HR fits to the measured SFI velocity field are
shown in figures 3 and 4, in redshift shells 2000\kms thick. The
infall to Virgo ($l
=284^\circ, b = 74^\circ$) dominates the
nearby SFI flow. Other features like Ursa Major and Fornax
are poorly sampled by the SFI. In the middle panel, the field exhibits
a dipole pattern corresponding to the reflex motion of the Local
Group with infalling galaxies in the Hydra-Centaurus direction and
an outward flow in the Perseus-Pisces direction, as seen in the LG
restframe, which is even stronger in the most distant shell.
Comparison of figures 3 and 4 show that despite some small amplitude
variations the LR and HR fits are remarkably similar over the entire
volume. The good agreement between the two fits
 is a consequence of the uniform sampling of SFI data. Our
ability to fit a higher resolution function also indicates that
the solution is stable and insensitive to the smoothing scale.

For comparison, we show in figure 5 the LR \iras field reconstructed
with $\beta=1$.  Comparing figures 3 and 5 one immediately sees
that although the amplitude of the \iras field with $\beta=1$ clearly
does not match that of the SFI field, the general pattern of the
velocity fields is remarkably similar with excellent agreement in
the locations of outflows and inflows. This result
builds confidence in the possiblity of determining an
accurate value of $\beta$ from the velocity-velocity comparison
using the observed SFI and predicted \iras fields. To illustrate
this, figure 6 shows the \iras flow with $\beta=0.6$ which clearly
yields a much better match to the amplitude of the SFI flow than that
obtained with $\beta =1$. The quality of the match can be evaluated in
figures 7 and 8, plotting the residuals from the comparison of
the SFI and \iras fields, with $\beta=0.6$ for the LR and HR fits,
respectively.  The overall agreement is remarkable with only a few
nearby galaxies giving large residuals. Most encouraging is the
absence of large regions of coherent residuals and the absence of
systematic residuals such as the dipole residual seen in previous
analysis at intermediate and distant redshift shells (DNW). Moreover,
the amplitude of the residuals is close to that obtained by DNW in the
analysis of mock catalogs.

\subsection {Determination of $\beta$}

The filtered SFI and \iras velocity fields are fully described by the
modal expansion coefficients, $\alpha^j_{tf}$ and $\alpha^j_{\iras}$.
Since the number of these coefficients is significantly smaller than
the number of galaxies, it is more efficient to estimate $\beta$ by
comparing the modes rather than the velocities of galaxies.  Following
DNW, we define our best estimate for $\beta$ as the value which
renders a minimum in the Pseudo $\chi^2$
\begin{equation}
\tilde \chi^2(\beta)=\sum_{j, j^\prime}\left[\alpha^j_{tf}-\alpha^j_{iras}(\beta)\right]
\left[{\rm \bf T}+ {\rm \bf M}(\beta)\right]^{-1}
\left[\alpha^{ j^\prime}_{tf}-\alpha^{j^\prime}_{iras}(\beta)\right] ,
\label{chia}
\end{equation}
where ${\rm \bf T\equiv}<\delta \alpha^j_{tf} \delta
\alpha^{j^\prime}_{tf}>$ 
and ${\rm \bf M}\equiv <\delta \alpha^j_{iras} \delta
\alpha^{j^\prime}_{iras}>$ are the the error covariance matrices of
the coefficients $\alpha^j_{tf}$ and $\alpha^j_{iras}$ respectively.
The ITF error matrix ${\rm {\bf T}}$ is easy to evaluate,
thanks to the orthonormality condition. This matrix is
diagonal and is given by
\begin{equation}
T^{jj^\prime} 
=\left(\frac{\sigma_{\eta}}{\gamma}\right)^2 \delta^K_{j{j^\prime}} .
\end{equation}

The \iras error covariance matrix is more cumbersome
to compute.  It should incorporate three sources of errors in the
\iras velocity field: $(i)$ the peculiar velocities are generated
 using galaxy redshifts relative to LG frame.  Any error in the LG motion
creates a dipole discrepancy between the SFI and the \iras
velocities. $(ii)$ the
\iras density field is estimated from a discrete distribution of
galaxies and therefore suffers from Poisson error which propagates
into the velocity field.  $(iii)$ small scale coherent (as in triple
valued zones) nonlinear velocities are not included in the \iras
recovered velocities and can be important in the error budget.  Note
that, in contrast to DNW, we have not included uncertainties due to
small scale incoherent (local velocity dispersion) velocities. That is
justified because the SFI modes suffer from a similar error
which roughly equals the error in the \iras modes. 
Moreover, since our estimated value for $\sigma_{\eta}$ includes
scatter due to incoherent velocities of galaxies in the SFI sample,
a somewhat lower value for $\sigma_{\eta}$ than estimated above
should be used in evaluating $\tilde \chi^2$. Below we will estimate
$\beta$ for various values of $\sigma_{\eta}$.

We can express the \iras velocity covariance as the sum of these
errors
\begin{equation}
<\Delta u_i \Delta u_j>=C_{LG}(i,j)+C_{SN}(i,j)+ C_{NL}(i,j) . \label{cuu}
\end{equation}
where the terms $C_{LG}$, $C_{SN}$ and $C_{NL}$
depend on $\beta$ and describe uncertainties due to LG motion, shot-noise and
nonlinearities, respectively.
  The modes covariance matrix,
${\rm \bf M}$, is then computed by projecting the elements of the
\iras velocity covariance matrix $<\Delta u_i
\Delta u_j>$ into the space of the base functions $F^j$ as
described in DNW (eq. 21).  In contrast to DNW, we argue here that the term,
$C_{LG}$, describing the effect of an error in the LG motion, depends
on $\beta$ for the following reason.  Both, the SFI and the
\iras fields suffer from errors resulting from the LG motion.  Since
only the difference $\alpha^j_{tf}-\alpha^j_{iras}$ enters in $\tilde
\chi^2$ in (\ref{chia}), we must take into account any cancelation of
the LG error in this difference.  In the SFI velocities, this error
amounts to a dipole term (filtered by the modal expansion).  The
effect in the \iras velocities is slightly more complicated.  An error
in the LG motion, generates an artificial dipole component in the
distribution of the \iras galaxies in redshift space which,
consequently, leads to a $\beta$ dependent velocity dipole in the
recovered \iras velocity field.  According to Nusser \& Davis (1994) a
dipole distribution generates a velocity dipole which, approximately,
scales like $\beta/(1+\beta)$.  That implies that the residual error
between the SFI and the \iras field scales like $1/(1+\beta)$.
Therefore we write
\begin{equation}
C_{LG}=\frac{\sigma_{LG}^2}{\left(1+\beta\right)^2} \cos(\theta_{ij}) 
\end{equation} 
where $\theta_{ij}$ is the angle between the lines of sight to
the points $i$ and $j$ and we set $\sigma_{LG}=150$\kms (see DNW).

We compute the Poisson error covariance matrix $C_{SN}$ by generating
18 bootstrap realizations of the observed 1.2 Jy \iras galaxy
distribution by replacing each observed galaxy with a number of points
drawn from a Poisson distribution with mean of unity. Then we compute
the velocity fields from these realizations by the same algorithm used
in the derivation of the velocity field from the observed galaxy
distribution.  For galaxy $i$ of the SFI sample, we tabulate the
differences, $\delta u_i$, between the velocity obtained from each of the
18 bootstrap realizations and the velocity as predicted from the
actual distribution of galaxies, and evaluate the covariance matrix
$C_{SN}$ by averaging the product $\delta u_i\delta u_j$ from all the
bootstrap realizations.  This process is computed for several
different values of $\beta$.

As in DNW, we model the non-linear term $C_{NL}(i,j)$ as
\begin{equation}
C_{NL}(i,j) =  \left((s_h(i) + s_h(j)\right)^2 
{\rm exp}\left(- {|\Delta {\bf s}(i,j)|^2 \over 2 \sigma_{coh}^2 } \right)  
\end{equation}
Where $\Delta {\bf s}(i,j)$ is the redshift separation between the
galaxies $i$ and $j$.  
We assume that the coherent error is
proportional to the square of the average shear in the
\iras derived velocity fields for each $\beta$,  
$s_h \equiv \sigma_{NL}~{\rm min}(1.67,|d v_p/dz|)$ \kms. We adopt a
value of $\sigma_{coh}=200$ \kms.  It is important to note that the
amplitude, $\sigma_{NL}$, of this error is uncertain. Here, we choose
$\sigma_{NL}=90$\kms, which makes the value of the reduced $\tilde \chi^2$ 
unity at the minimum for $\sigma_{\eta}=0.065$. The choice is natural based on
the hypothesis that the SFI and \iras fields independently describe
the same underlying velocity field. Moreover the amplitude of the
error is reasonable. However, we will compute $\tilde \chi^2(\beta)$
for various values of this error. Fortunately, the best $\beta$
estimate is robust with repect to changes in the amplitude of this
error.

Given the covariance matrices, we compute curves of the reduced $\tilde
\chi^2(\beta)$, for the HR and LR filters.  The results are summarized
in figure 9, which shows curves of the reduced $\tilde \chi^2(\beta)$
for the two filters using various error estimates, as explained in the
figure caption.  For both filters, the minimum value of the $\tilde
\chi^2$ is attained at $\beta=0.6$ regardless of the details of our
estimate of the covariance matrices. The 1-sigma error is less than
$0.1$ for all the curves.  We have also computed curves of the reduced $\tilde
\chi^2(\beta)$ for values of $C_{LG}$ and $C_{NL}$ not shown
in figure~9 and find that the minimum is not sensitive to their
exact values.  The amplitude of $C_{NL}$ was chosen so that the value
of $\tilde \chi^2$ per degree of freedom for the LR filter is very
close to unity when the nonlinear error is included.  The
corresponding value for the HR filter is 1.23, significantly larger
than unity. This large value possibly indicates some disagreement
between the fields on smaller scales. However, it is reassuring that
the visual inspection of the fields reveals no gross
discrepancies between them (figure~8). Note also that a large value of
$\tilde \chi^2$ can be attributed to the fact that the scatter in the
ITF relation is not strictly Gaussian, and since the effects of
non-gaussianity are more important on small scales, we expect a larger
deviation from a value of unity of the reduced $\tilde
\chi^2$ for the HR filter. To evaluate the effect of the
$\beta$ dependence in the term $C_{LG}$, we computed $\tilde \chi^2$
using the expression of DNW which did not include the factor
$1/(1+\beta)$. Although the shape of the curve became flatter for
$\beta>0.6$ and steeper for smaller values of $\beta$, the minimum remained
unchanged at $\beta=0.6$.

It is interesting to inspect the correlation function of the residual
fields. Figure 10 plots the correlation function of the quantity
$P_{SFI}-P_{IRAS}$ for various values of $\beta$, where
$P=-5\log(1-u/s)$
as defined in section 3.  For comparison,
the correlation function of $P_{SFI}$ is also shown.  The overall
amplitude of the correlation function corresponding to the residual
field for $\beta=0.4$ or $0.6$ is significantly smaller than that of
the SFI field.  Note that the curve corresponding to $\beta=0.6$ which
minimizes $\tilde \chi^2$ does not have the lowest amplitude at zero
lag. The reason is clearly that the value of the correlation function
at zero lag is simply the variance of the residual field and cannot be
used to determined the best fit $\beta$ as it does not include the
covariance of the errors.

\void{
\begin{center}
{\bf TABLE 1: Coefficient Comparison} \\
\begin{tabular}{|c|c|c|} \hline
$\beta$ & \multicolumn{2}{c|}{${\tilde \chi}^2$} \\ \hline 0.2 & 72.70
& 133.5\\ \hline 0.4 & 47.87 & 103.6\\ \hline 0.5 & 42.71 & 97.24 \\
\hline 0.6 & 40.33 & 94.63 \\ \hline 0.7 & 41.35 & 95.29 \\ \hline 0.8
& 44.03 & 98.46 \\ \hline 1.0 & 55.10 & 111.9 \\ \hline
\end{tabular}
\end{center}
\begin{center}
{\bf TABLE 1: Table of $\tilde \chi^2(\beta)$} \\
\vspace*{5mm}
\begin{tabular}{|c|c|c|c|c|} \hline
 & \multicolumn{2}{c|}{with $C_{NL}$} & 
\multicolumn{2}{c|}{without $C_{NL}$ } \\ \cline{2-5}
$\beta$&LR&HR&LR&HR \\ \hline
0.2 & 68.77 & 123.1 &72.74&128.5 \\ \hline
0.4 & 46.84 & 97.96 &50.81&104.1 \\ \hline
0.5 & 42.26 & 92.87 &46.24&99.44 \\ \hline
0.6 & 40.41 & 91.33 &44.76&98.63 \\ \hline
0.7 & 41.64 & 93.10 &46.06&101.27 \\ \hline
0.8 & 44.69 & 97.52 &49.64&107.1 \\ \hline
1.0 & 56.96 & 114.6 &63.98&130.3 \\ \hline
\end{tabular}
\end{center}
}

\subsection{Comparison between the SFI and Mark~III Results}

As pointed out in the introduction, the ITF method has recently been
applied to the Mark~III catalog by DNW and their results can be used
for an indirect assessment of the differences between the SFI and
Mark~III catalogs.  DNW used 2900 spirals,
in the Mark~III catalog, including cluster and field galaxies.
Notwithstanding the larger number of galaxies in the Mark~III sample,
the non-uniform sampling of the surveyed volume prevented DNW from
carrying out the modal expansion with more than 56 modes of an effective
resolution intermediate between our LR and HR fits.

Comparing our figure 3 with figure 8 of DNW we see that nearby the
Mark~III has considerably more points as it includes the Aaroson \etal
(1982) data.  In the intermediated redshift shell the most significant
difference is that the strong velocity gradient seen in the Mark~III
in the interval $240^\circ < l < 330^\circ$ and $15^\circ < b <
40^\circ $ is less pronounced in the SFI data. The flow there is
predominantly determined by the MFB92 data which have been pruned to
include only Sbc-Sc galaxies in the SFI (see section 2).  In the SFI
map there is an infall motion at ($210^\circ< l< 270^\circ, -40^\circ
<b<-20^\circ$) which is absent in the Mark~III flow.  This infall
extends smoothly northward to the Great Attractor region.  The most
obvious differences between Mark~III and SFI, not surprisingly, are
seen in the last redshift shell.  The general impression is that the
dipole patterns in the two maps differ significantly in direction and
amplitude.  The direction of the dipole in the SFI map seems to lie
along the direction of the reflex motion of the LG, while the Mark~III
dipole is directed along the south galactic pole.  In sharp contrast
to the SFI velocity field, the Mark~III flow exhibits a strong infall
in the northern cap. In the general direction of PP (lower
left part of the slice), both SFI and Mark~III flows have the same
sign but with slightly different amplitudes. Note that the SFI
galaxies cover that region more uniformly, illustrating the overall
uniformity of the SFI.

\section{Summary and Discussion}

The principal advantage of estimating $\beta$ from velocity-velocity and
density-density comparisons is that they are model independent. This
is in contrast to other methods like the power spectrum analysis
(Zaroubi \etal 1996) and redshift distortions (\eg Fisher \etal 1994,
Hamilton 1993).  Although the velocity-velocity and density-density
comparisons are equivalent, in practice, they differ in several
aspects.  In particular velocity-velocity comparison on the basis of
the modal expansion done here, requires no corrections to the
biases which plague density-density comparisons. Moreover, nonlinear
effects are less significant in the velocity-velocity comparison.

We have compared the observed SFI peculiar
velocity field with that predicted from the galaxy distribution of the
\iras galaxies.  We have found good agreement between the two fields
for $\beta = 0.6 \pm 0.1$ with no dependence on depth. Similar values
for $\beta$ have been obtained by Freudling \etal (1997).  We have
determined $\beta$ using filters with two different resolutions and
found no evidence for scale dependence in the estimate of $\beta$.
The good agreement between the fields is consistent with the galaxy
distribution being closely related to that of the dark matter by means
of a scale independent biasing factor, and with the hypothesis that
the TF relation does not depend strongly on environment.  It seems
unreasonable that the biasing mechanism and the TF relation should
conspire to yield such a good agreement between the fields.

Our estimate for $\beta$ is consistent with the value determined by
DNW from the comparison of the Mark~III with the \iras field. However,
DNW found systematic discrepancies between the fields, which prevented
a firm determination of $\beta$ from their analysis. The major problem
was the presence of a residual dipole component which strongly
increased with depth beyond $3000$\kms. Our residual fields show no
persisting coherent features in the comparison volume. That gives us
confidence in our estimate for $\beta$ and that the SFI flow field is
a fair representation of the underlying velocity field.

The Mark~III and the SFI flows seem exhibit some disagreement beyond a
redshift of $3000$\kms.  The differences between the flows at lower
redshifts are not serious and can be understood in terms of the
different selection criteria used in defining the two samples.  Willick
\etal (1997) presented a likelihood analysis of the \iras and Mark~III
velocity fields within a distance of 3000 \kms. They found a good
match between the Mark~III and the \iras fields in that limited volume
and derived a value of $\beta \sim 0.5$. This is consistent with our
estimate of $\beta$ obtained from the analysis of the peculiar
velocity field within 6000 \kms. This is another indication that the
differences between the SFI and the Mark~III catalog are to be found primarily at
large distances. Resolving these discrepancies should allow us to
combine the two samples in order to
better constrain the local peculiar velocity field
which is of great interest.

Estimates of $\beta$ from velocity-velocity comparisons point
towards values of $\beta$ lower than obtained previously with
different samples and other methods (\eg Dekel \etal 1993, Zaroubi
\etal 1996).  A value of $\beta$ = 0.6 is consistent with other
determinations obtained independently of large scale flows, \eg,
cluster abundance, galaxy power-spectrum and small-scale velocities.
If the universe is flat, a value $\beta\sim 0.6$ requires a biasing
factor of $\sim 2$ for optical galaxies. Such a high value for the
biasing value implies a rather low normalization for the matter
density fluctuations which is hard to reconcile with COBE-normalized
CDM-like models. Moreover, such a large value for $b$ is difficult to
obtain in hierarchical structure formation (Kauffman, Nusser and
Steinmetz 1997).

The results presented here are encouraging and demonstrate the
importance of all-sky homogeneous measurements of peculiar velocities.
Current samples are still too small and sparse to allow them to
be used to explore smaller scales ($ 500-1500$ \kms) where non-linear
effects become important. Future samples should aim at a higher
sampling rate and uniform sky coverage.

\section*{ Acknowledgements}
We would like to thank Marc Davis for many valuable discussions and
suggestions. The results presented in this paper are based on
observations carried out at the National Astronomy and Ionosphere
Center (NAIC), the National Radio Astronomy Observatory (NRAO), the
Kitt Peak National Observatory (KPNO), the Cerro Tololo Interamerican
Observatory (CTIO), European Southern Observatory, the Palomar
Observatory (PO), the Observatory of Paris at Nan\c cay and the
Michigan--Dartmouth--MIT Observatory (MDM). NAIC, NRAO, KPNO and CTIO
are respectively operated by Cornell University, Associated
Universities, inc., and Associated Universities for Research in
Astronomy, all under cooperative agreements with the National Science
Foundation. Access to the 5m telescope at PO was guaranteed under an
agreement between Cornell University and the California Institute of
Technology.  This research was supported by NSF grants AST94--20505 to
RG, AST90-14850 and AST90-23450 to MH and AST93--47714 to GW who would also
like to acknowledge support by ESO for a visit to Garching.

\clearpage


\begin{figure}
\centering
\mbox{\psfig{figure=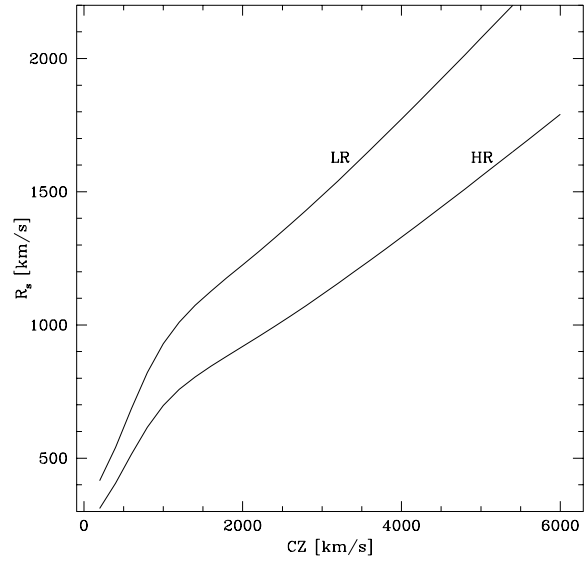,height=8cm,bbllx=18pt,bblly=144pt,bburx=592pt,bbury=718pt}}
\caption{
The radial resolution scale of the filters HR and LR as a function of 
redshift. At any redshift, the smoothing of the filters roughly corresponds
to a sharp cutoff of $k_s=2\pi/R_s$ in k-space. } 
\end{figure}

\clearpage

\begin{figure}
\centering
\mbox{\psfig{figure=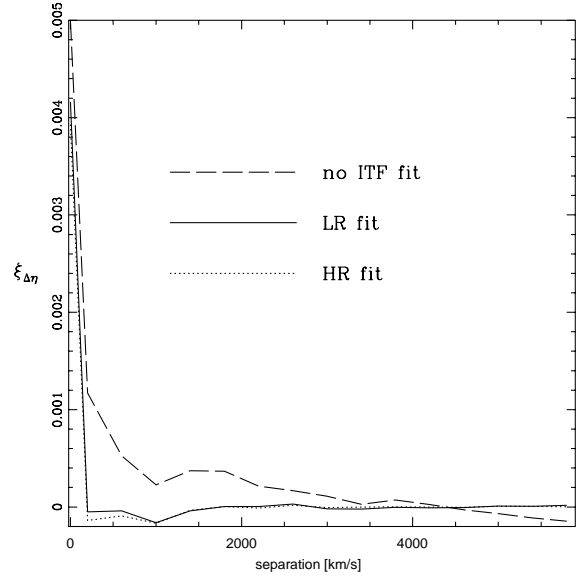,height=8cm,bbllx=18pt,bblly=144pt,bburx=592pt,bbury=718pt}}
\caption{
The autocorrelation function of the $\eta$ residuals of the
SFI galaxies versus
redshift space separation.
The dashed curve is before the ITF fitting, while the solid and dotted
are after the ITF fitting with the LR and HR respectively.}
\end{figure}

\clearpage

\begin{figure}
\centering
\mbox{\psfig{figure=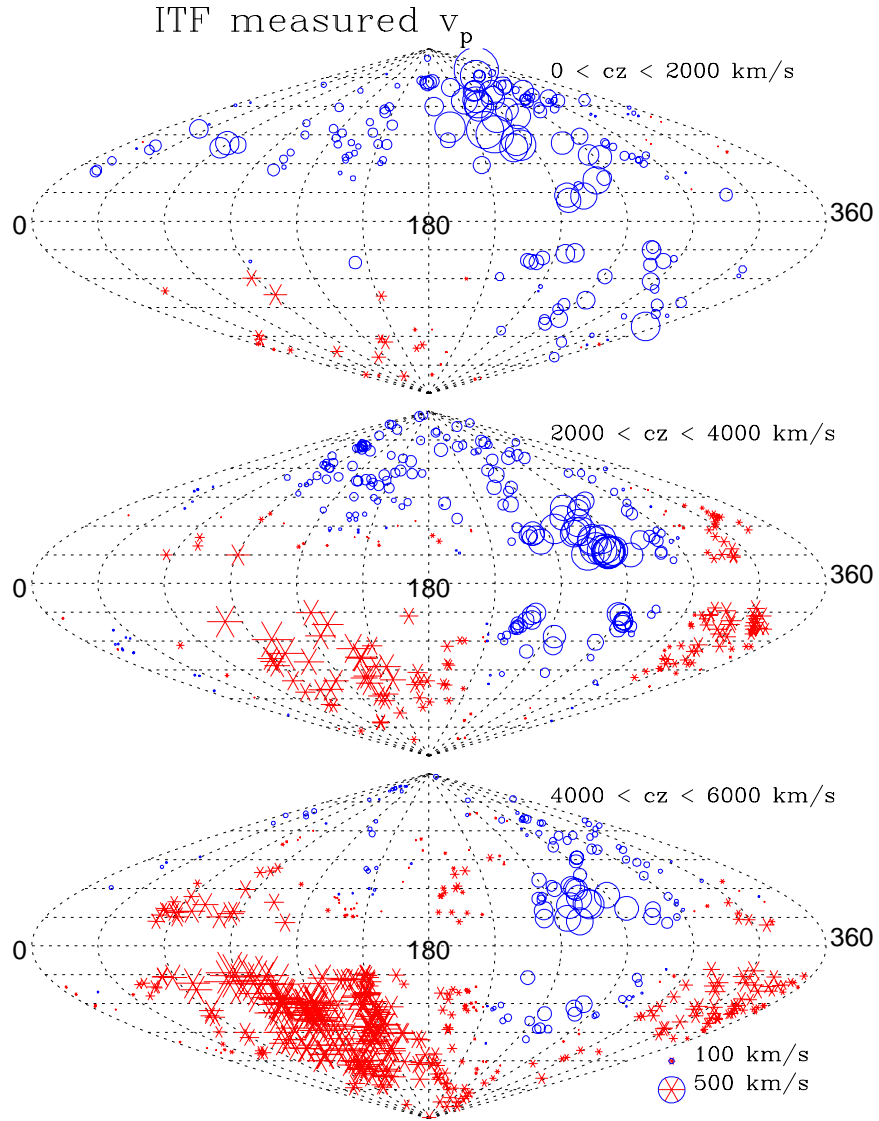,height=15cm,bbllx=18pt,bblly=144pt,bburx=592pt,bbury=718pt}}
\caption{The sky projection in galactic coordinates  as seen in the
 LG frame of the low resolution ITF velocity field for the SFI galaxies 
}
\end{figure}

\clearpage

\begin{figure}
\centering
\mbox{\psfig{figure=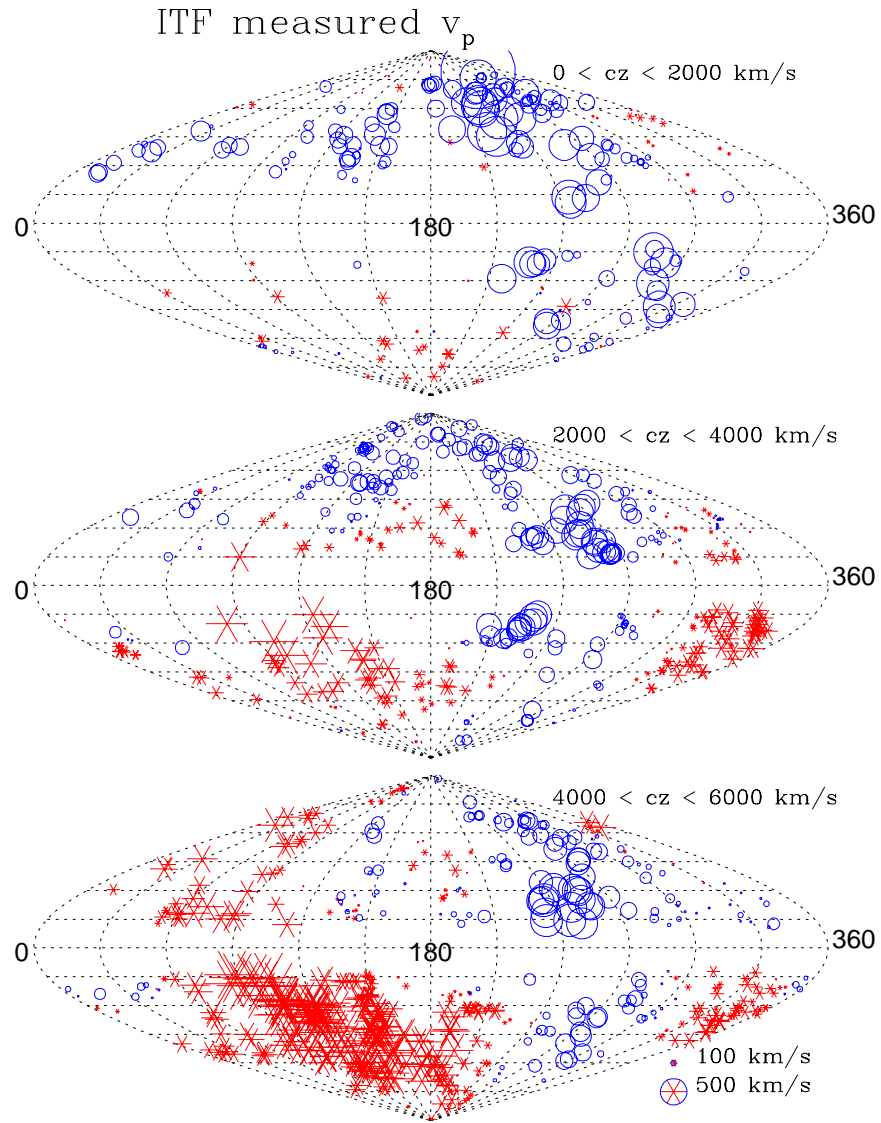,height=15cm,bbllx=18pt,bblly=144pt,bburx=592pt,bbury=718pt}}
\caption{The high resolution ITF velocity field for the SFI galaxies
using the same sky projection as figure 3
}
\end{figure}

\clearpage

\begin{figure}
\centering
\mbox{\psfig{figure=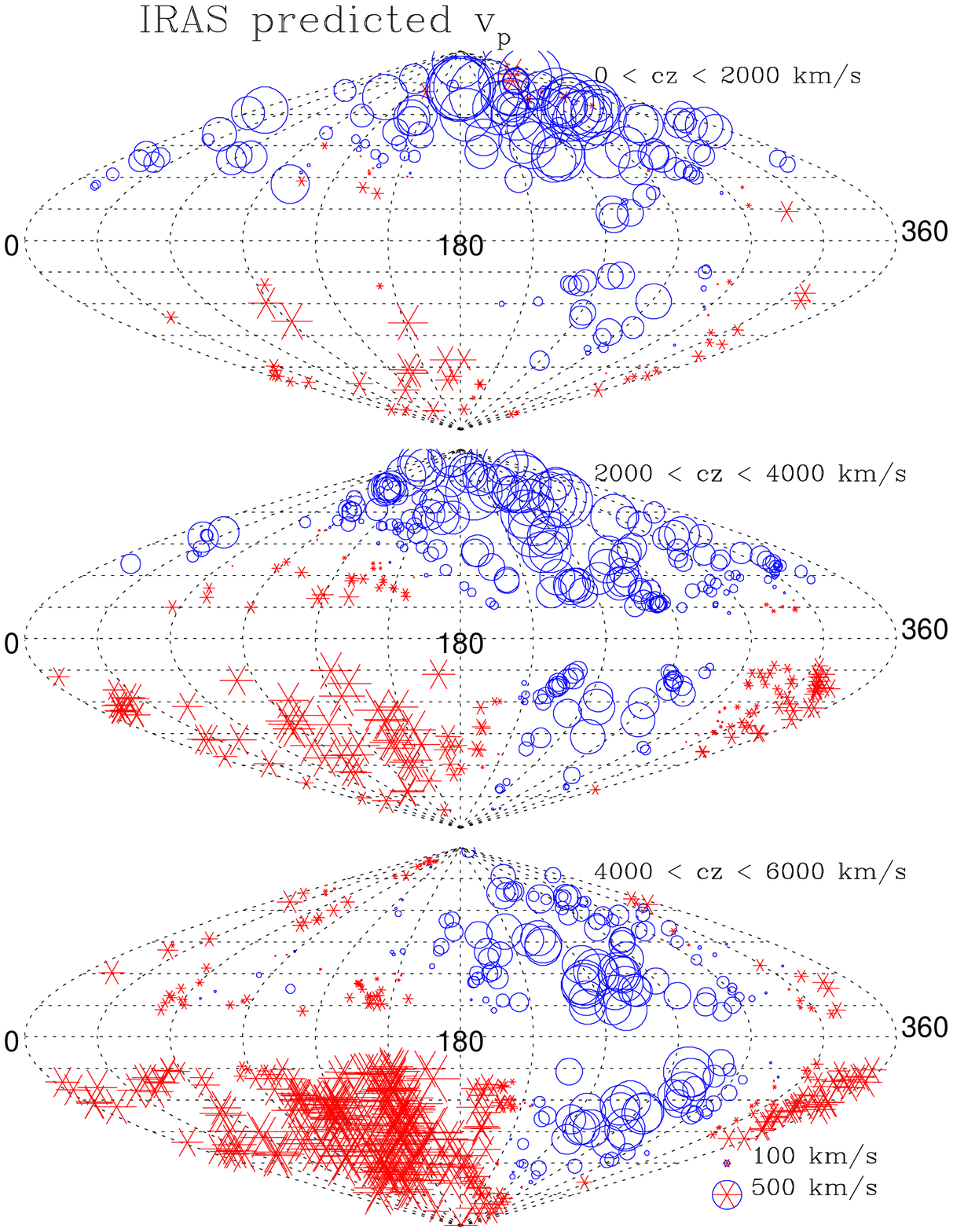,height=15cm,bbllx=18pt,bblly=144pt,bburx=592pt,bbury=718pt}}
\caption{The low resolution  velocity field for the {\it IRAS}  galaxies
for $\beta =1$ }
\end{figure}

\clearpage

\begin{figure}
\centering
\mbox{\psfig{figure=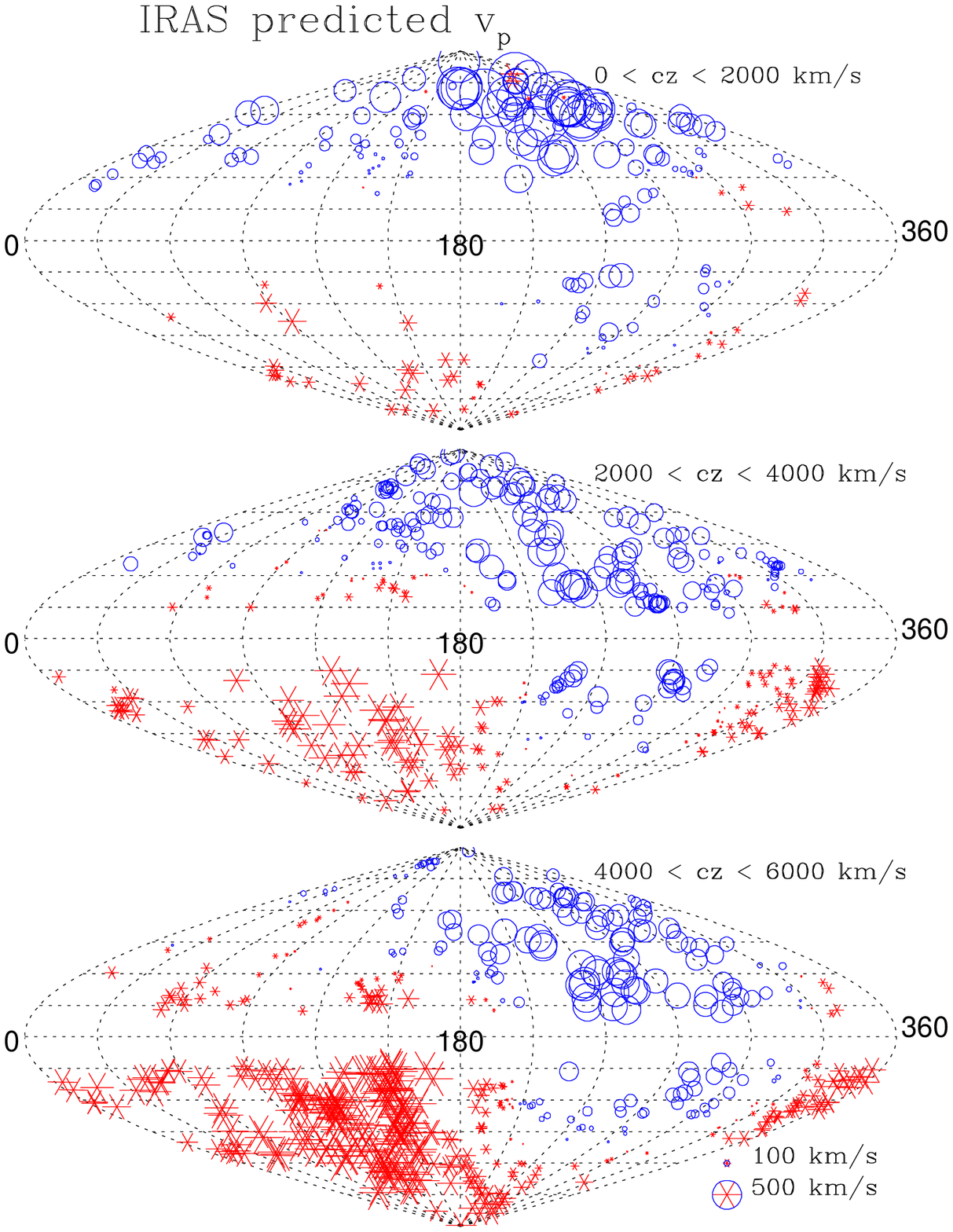,height=15cm,bbllx=18pt,bblly=144pt,bburx=592pt,bbury=718pt}}
\caption{The low resolution  velocity field for the {\it IRAS}  galaxies
for $\beta =0.6$ }
\end{figure}

\clearpage

\begin{figure}
\centering
\mbox{\psfig{figure=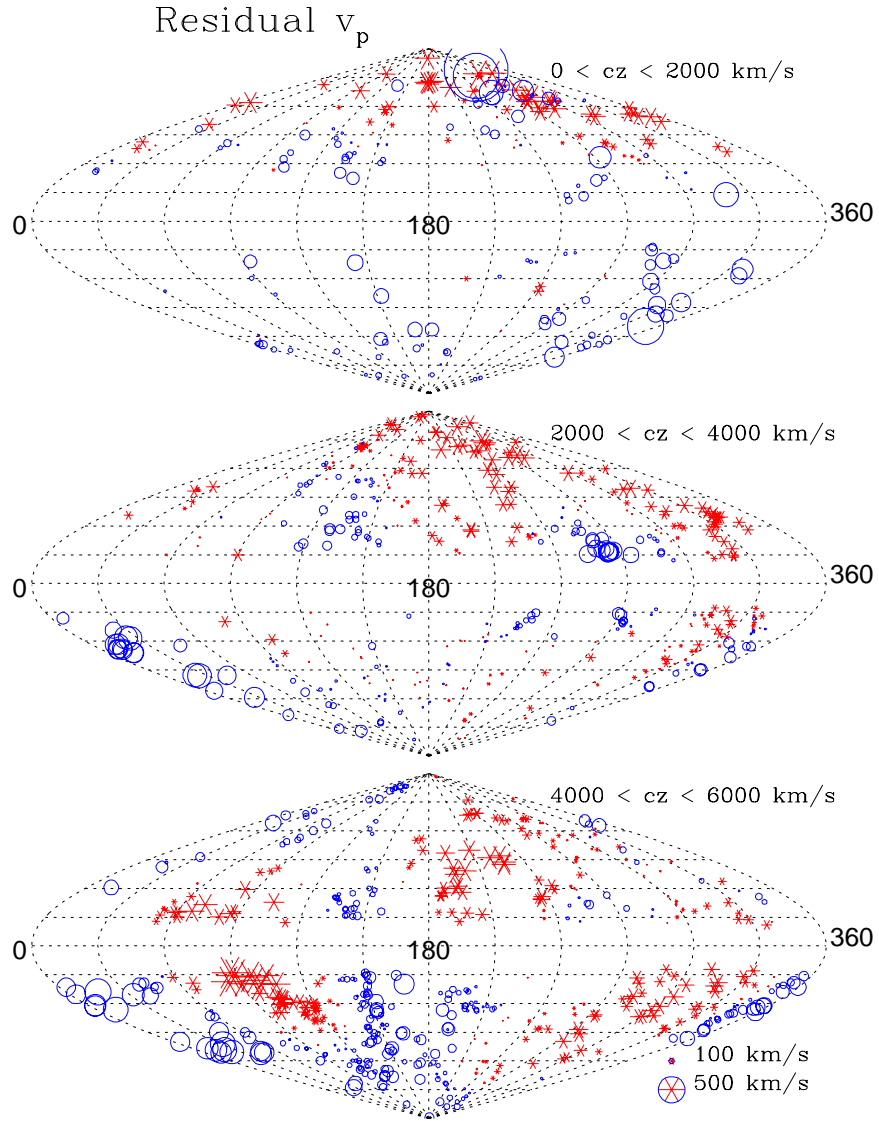,height=15cm,bbllx=18pt,bblly=144pt,bburx=592pt,bbury=718pt}}
\caption{The sky projection of the residuals $u_{SFI} -u_{{\it IRAS}}$
for $\beta=0.6$, for  low resolution fits}
\end{figure}

\clearpage

\begin{figure}
\centering
\mbox{\psfig{figure=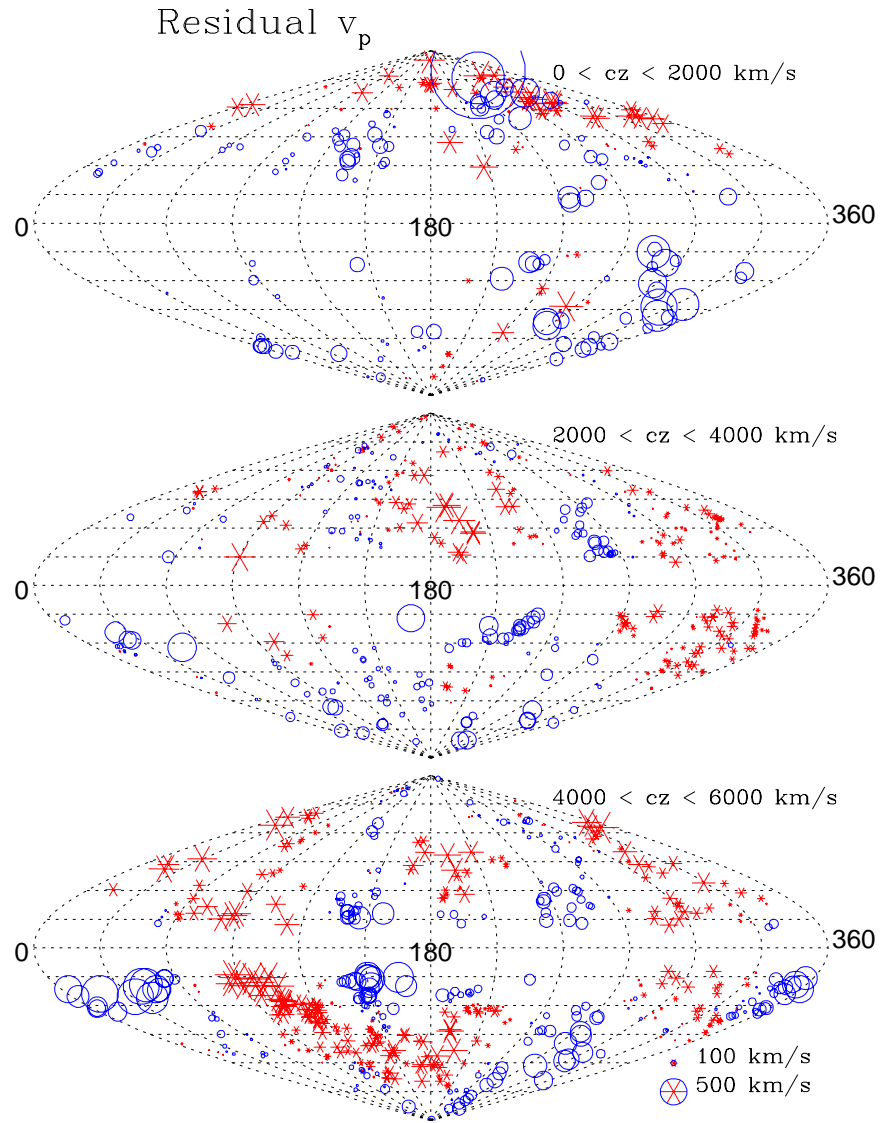,height=15cm,bbllx=18pt,bblly=144pt,bburx=592pt,bbury=718pt}}
\caption{The sky projection of the residuals $u_{SFI} -u_{{\it IRAS}}$
for $\beta=0.6$, for the high  resolution fits}
\end{figure}

\clearpage

\begin{figure}
\centering
\mbox{\psfig{figure=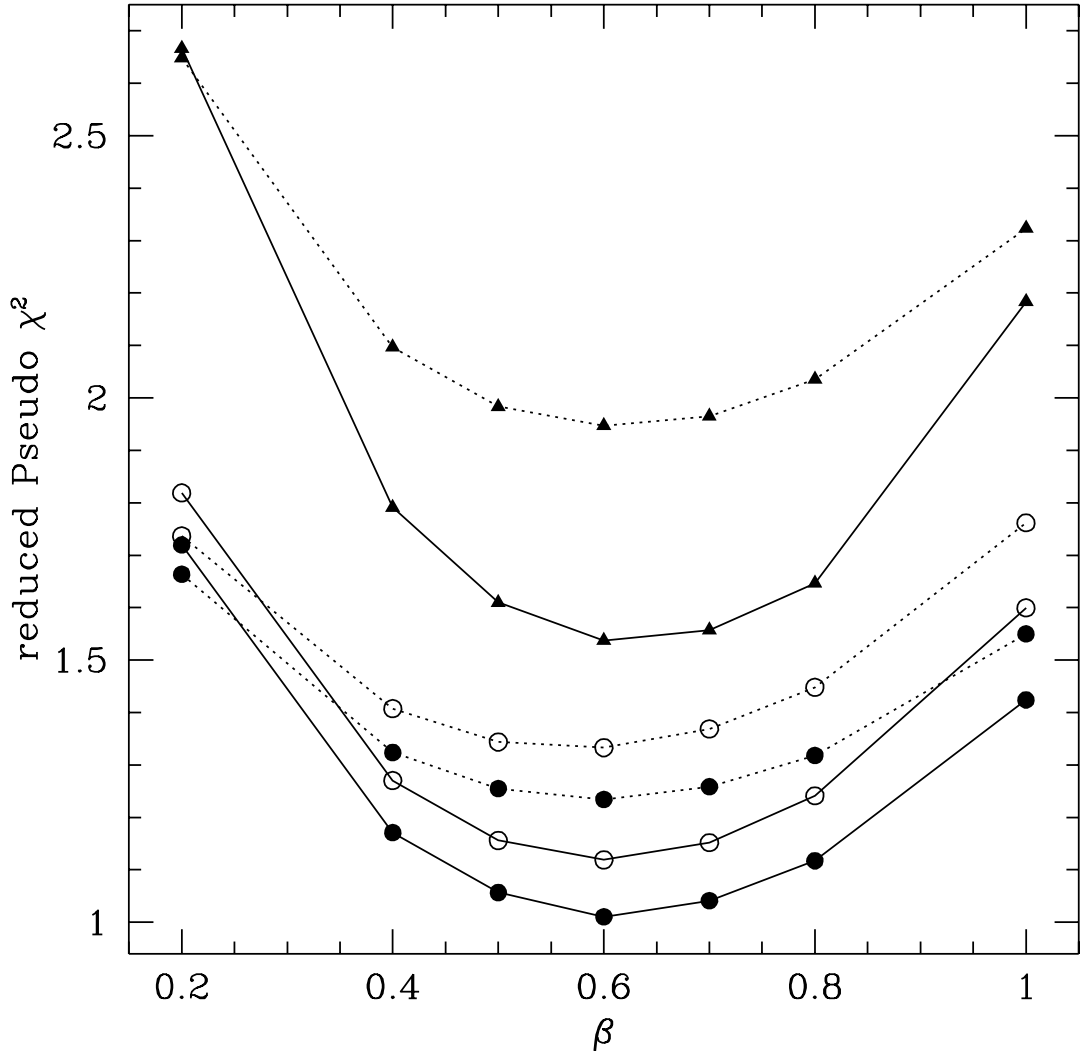,height=15cm,bbllx=18pt,bblly=144pt,bburx=592pt,bbury=718pt}}
\caption{Curves of reduced Pseudo-$\chi^2$ verus $\beta$ computed
using equation (6) for the LR (solid lines) and HR (dotted lines),
filters. Shown are curves for two different values for $\sigma_{\eta}$
= 0.065 (circles) and 0.05 (triangles), with (filled symbols) or
without (open symbols) the non-linear error $C_{NL}$.}
\end{figure}

\clearpage

\begin{figure}
\centering
\mbox{\psfig{figure=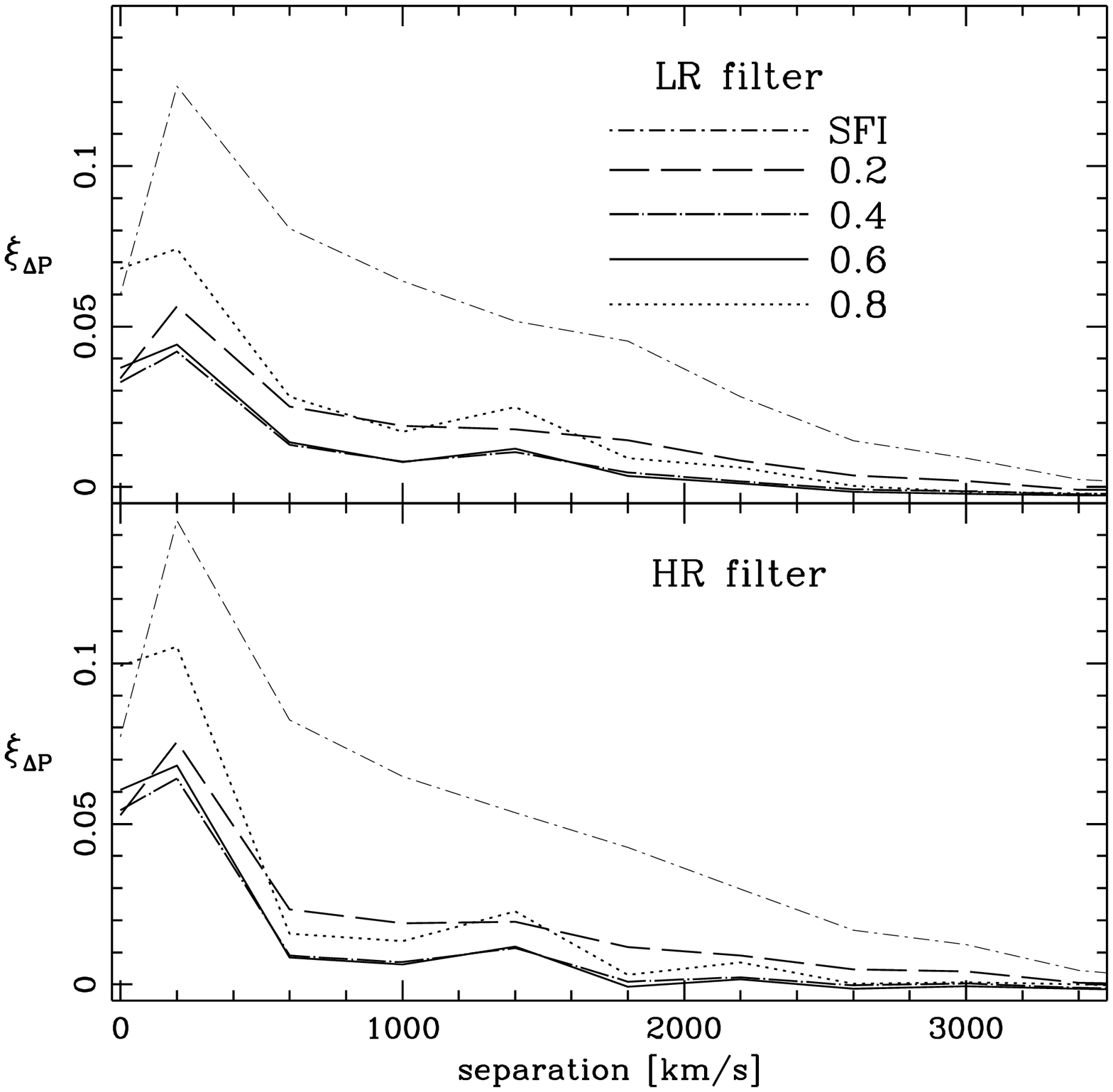,height=15cm,bbllx=18pt,bblly=144pt,bburx=592pt,bbury=718pt}}
\caption{The correlations function of $\Delta P = P_{SFI} - P_{IRAS}$
for various values of $\beta$ as inidicated in the plot. For
comparison we also show the correlation function of the SFI field}
\end{figure}


\begin{thebibliography}{}
\bibitem{aa86} 
Aaronson, M.,  Huchra, J., Mould, J., Schechter, P. \& Tully, R.B.,
1982, ApJ ~258, 64

\bibitem{}
Carlberg, R.G., Yee, H.K.C., Ellingson, E., Abraham, R., Cravel, P., Morris,
S. \& Pritchet, C.J. 1996, ApJ 462, 32

\bibitem{dac96}
da Costa, L.N., Freudling, W., Wegner, G., Giovanelli, R., Haynes, M.P. \&
Salzer, J.J. 1996, ApJ, 468, L5

\bibitem{}
Davis, M., Nusser, A. \& Willick, J., 1996, ApJ, 473, 22 (DNW)

\bibitem{}
Dekel, A., Bertschinger, E. Yahil, A., Strauss, M, Davis, M. \& Huchra, J.,
1993, ApJ 412, 1




\bibitem{fish}
Fisher, K. B. , Davis, M., Strauss, M., Yahil, A. \& Huchra, J., 1994,
MNRAS 267, 927

\bibitem{}
Fisher, K.B., Lahav, O., Hoffman, Y., Lynden-Bell, D. \& Zaroubi, S., 1995,
MNRAS 272, 885


\bibitem{fre95}
Freudling, W.,da Costa, L.N.,  Giovanelli, R., Haynes, M.P., Salzer, J.J.
\& Wegner  1997, {\it in preparation}



\bibitem{gio96} 
Giovanelli, R., Haynes, M.P., Herter, T., Vogt, N.,
Salzer, J.J., Wegner, G., da Costa, L.N. \& Freudling, W. 1997a, AJ ~113, 22

\bibitem{gio96} 
Giovanelli, R., Haynes, M.P., Herter, T., Vogt, N.,
da Costa, L.N., Freudling, W. Salzer, J.J. \& Wegner, G., 
1997b, AJ ~113, 53


\bibitem{}
Hamilton, A.J.S. 1993, ApJ, 406, 247

\bibitem{}
Haynes \etal 1997, {\it in preparation}

\bibitem{}
Kauffman, G., Nusser, A. \& Steinmetz, M. , 1997, MNRAS 286, 795

\bibitem{}
Kochanek, C.S., 1996, ApJ, 466, 638

\bibitem{mat93}
Mathewson, D.S., Ford, V.L. and Buchhorn, M. 1992, ApJS ~81, 413  (MFB92)

\bibitem{}
Nusser, A. \& Davis, M., 1994, ApJ, 421, L1

\bibitem{}
Nusser, A. \& Davis, M., 1995, ApJ, 449, 439 (ND95)


\bibitem{}
Perlmutter \etal, 1997, ApJ ~483, 565

\bibitem{per95} 
Persic, M. and Salucci, P. 1995, ApJS 99, 501

\bibitem{}
Schechter, P. 1980, ApJ 85, 801

\bibitem{}
Willick, J.A., Courteau, S., Faber S.M., Burstein, D., Dekel, A. and
Strauss, M. A. 1997, ApJS 109, 333

\bibitem{}
Willick, J.A., Strauss, M.A., Dekel, A \& Kollat, T., 1997, {\it preprint}

\bibitem{}
Zaroubi, S., Zehavi, I., Dekel, A. \& Kollat, T. 1997, ApJ. {\it in press}

\end{thebibliography}
\end{document}